\documentclass[journal,10pt,letterpaper,final,twocolumn]{IEEEtran}%
\usepackage{amsmath}
\usepackage{amsfonts}
\usepackage{cite}
\usepackage{amssymb}
\usepackage{mathrsfs}
\usepackage{graphicx}
\usepackage[usenames,dvipsnames]{color}
\usepackage{epstopdf}
\usepackage[all]{xy}
\usepackage{supertabular}
\usepackage{subcaption}
\usepackage{array}
\usepackage{xcolor}

\providecommand{\U}[1]{\protect\rule{.1in}{.1in}}
\pdfoutput=1
\providecommand{\U}[1]{\protect\rule{.1in}{.1in}}

\newcommand{\qed}{\nobreak \ifvmode \relax \else
      \ifdim\lastskip<1.5em \hskip-\lastskip
      \hskip1.5em plus0em minus0.5em \fi \nobreak
      \vrule height0.75em width0.5em depth0.25em\fi}

\begin{document}

\title{What Should 6G Be?}
\author{Shuping Dang, Osama Amin, Basem Shihada, Mohamed-Slim Alouini$^{*}$\\
Computer, Electrical and Mathematical Sciences and Engineering Division\\ King Abdullah University of Science and Technology (KAUST), 
Thuwal 23955-6900, Saudi Arabia \\e-mail: \{shuping.dang, osama.amin, basem.shihada, slim.alouini\}@kaust.edu.sa\\Tel:  +966 12 808-0283; 
Fax: +966 12 802-0143}
\twocolumn[
  \begin{@twocolumnfalse}
\maketitle

\begin{abstract}
The standardization of fifth generation (5G) communications has been completed, and the 5G network should be commercially launched in 2020. As a result, the visioning and planning of sixth generation (6G) communications has begun, with an aim to provide communication services for the future demands of the 2030s. Here we provide a vision for 6G that could serve a research guide in the post-5G era. We suggest that human-centric mobile communications will still be the most important application of 6G and the 6G network should be human centric. Thus, high security, secrecy, and privacy should be key features of 6G and should be given particular attention by the wireless research community. To support this vision, we provide a systematic framework in which potential application scenarios of 6G are anticipated and subdivided. We subsequently define key potential features of 6G and discuss the required communication technologies. We also explore the issues beyond communication technologies that could hamper research and deployment of 6G.
\end{abstract}
  \end{@twocolumnfalse}
]


\IEEEPARstart{S}{ince} the initial development of the Advanced Mobile Phone System (AMPS) by Bell Labs, which was later called the first generation (1G) network, there have been three large-scale and radical updates to wireless communication networks over the past four decades, resulting in the second, third, and fourth generation (2G, 3G, and 4G) networks  \cite{alsharif2017evolution}. The launch of the fifth generation (5G) network is ongoing and is expected to be commercialized by 2020. As the standardization of 5G has gradually been solidified, researchers have begun to consider the future sixth generation (6G) communication network \cite{8412482,8603730,8631208,8869705,8792135,tariq2019speculative}. 

In this Perspective, we consider what 6G should be. We believe that conventional mobile communications will still be the most important application of 6G around 2030, though other application scenarios will become ubiquitous and increasingly significant. Consequently, the 6G network should be \textit{human centric}, rather than machine centric, application centric, or data centric. Following this rationale, high security, secrecy, and privacy should be the key features of 6G. Furthermore, user experience (UE) would be adopted as a pivotal metric in 6G communication networks. To support this vision for 6G communications, we provide a comprehensive and systematic framework. Specifically, we first anticipate and subdivide the potential application scenarios of 6G. We then define key features of 6G and discuss the required enabling communication technologies. We also explore issues beyond the communication technologies that could significantly affect the research and deployment of 6G in the 2030s.

\section*{\textbf{Background}}
\begin{figure*}[!t]
\centering
\includegraphics[width=7.0in]{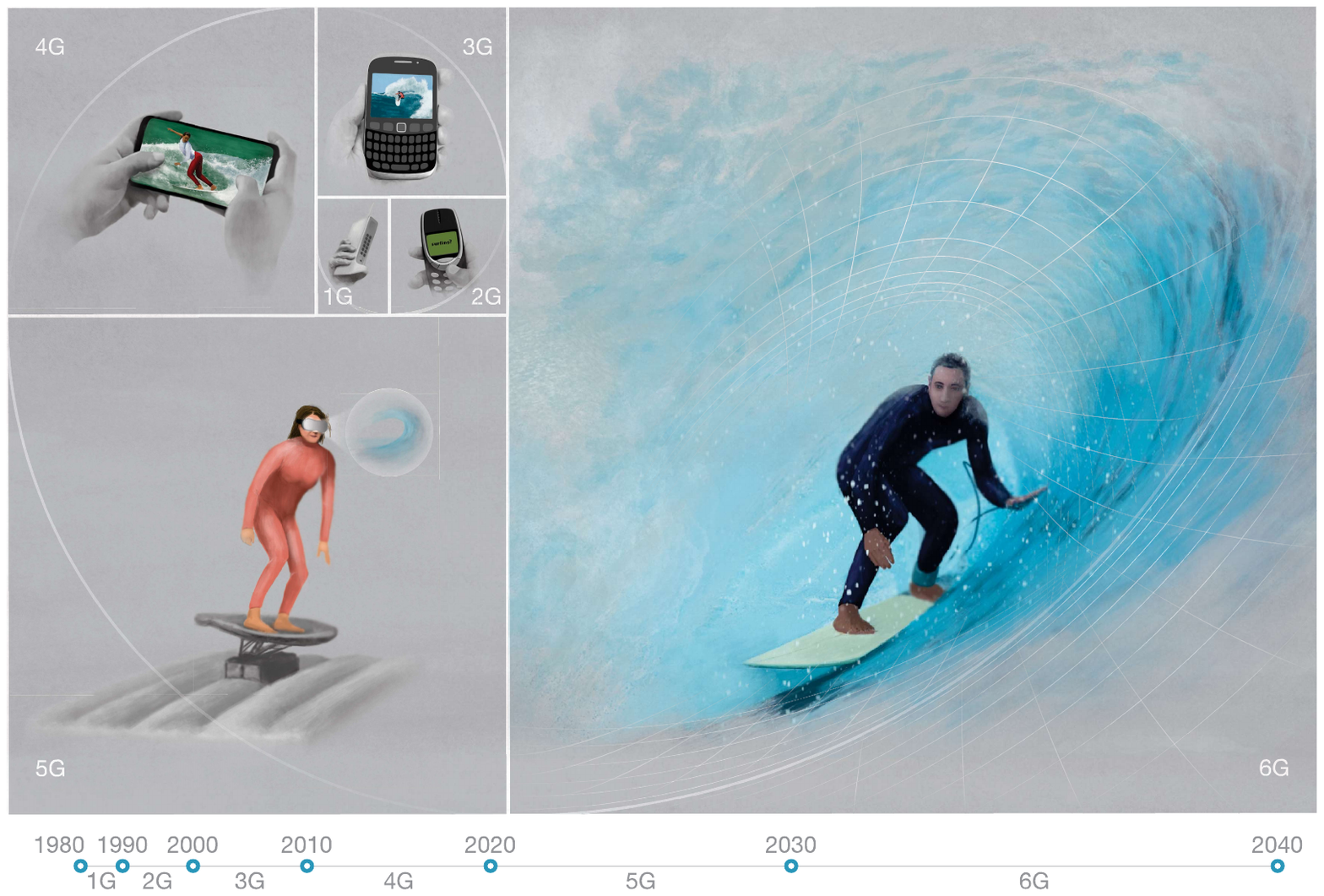}
\caption{\textbf{A user's perception of the different communications networks, from 1G to the hypothetical 6G.} In 1G and 2G, voice and text are available. In 3G and 4G, picture and video become commonplace. In 5G, live ultra-high-definition three-dimensional data can be employed. In 6G, it is expected that we could have a ubiquitous virtual existence.}

\label{evolution}
\end{figure*}

To justify our 6G vision, we first provide some background that covers network evolution from 1G to 4G, the 5G status quo, and the current research progress towards 6G (Fig. \ref{evolution}).

\subsection*{\textbf{Network Evolution from 1G to 4G}} Wireless communication stems from Marconi's pioneering demonstration of wireless telegraphy in the 19th century and was theoretically constructed based on information theory formed by Shannon in 1948. In the 1980s, the 1G analogue wireless cellular network was in use to allow mobile communications of voice, which was then replaced by the 2G digital cellular network in the early 1990s. Because of digitalization, 2G was capable of providing encrypted services and data services in addition to the traditional voice services, such as short messaging service (SMS). In the early 21st century, 3G, represented by wideband CDMA (WCDMA), CDMA2000, time-division synchronous CDMA (TD-SCDMA), and Worldwide Interoperability for Microwave Access (WiMAX), enabled various data services, including Internet access, video calls, and mobile television \cite{7000985}. In 4G/Long-Term Evolution (LTE) networks initialized in 2009, multiple-input and multiple-output (MIMO) antenna architecture, orthogonal frequency-division multiplexing (OFDM), and all-Internet protocol (IP) technology were jointly applied to achieve high-speed mobile data transmission \cite{6769024}. 4G has been a significant success both technological and commercially. With the proliferation of smartphones and tablets, mobile communications have become mainstream, providing a considerable amount of data throughput in 4G networks \cite{8412482}, and the information and communications technologies (ICTs) accompanying 4G have helped reshape society \cite{923566,4468730,Frias2012,moral2011technoeconomic,8750780}.

\subsection*{\textbf{What 5G Has Been}}
In 2014, a paper was published that discussed what 5G will be and pointed out that the key technologies to achieve 5G were network densification, millimetre wave, and massive MIMO architecture \cite{6824752}. Since then, the concept of 5G has been gradually solidified, and the main technological companies and operators have now launched their construction plans for 5G networks in order to deliver large-scale commercial deployment by 2020.

In the first deployment stage of 5G networks, most operators and device manufacturers adopt the 3GPP 5G New Radio (NR) standard for dense urban areas \cite{8258595}. The corresponding 5G network operates on the 2-6 GHz spectra. Both millimetre wave and massive MIMO technologies are widely used in 5G networks, while the network densification construction is delayed for certain reasons. Network slicing is more or less involved in 5G mission-critical solutions. Internet Protocol television (IPTV) and high-definition (HD) video streaming, service over high-speed mobility, basic virtual reality (VR) and augmented reality (AR) services can be well supported. Indoor services and data services in dense metropolitan areas will continue to be the main focus in the 5G era. For different application scenarios, a complete 5G communication network provides three service options: enhanced mobile broadband (eMBB), ultra-reliable low-latency communications (URLLC), and massive machine-type communications
(mMTC) \cite{8638891}. 

On the other hand, there are also a variety of state-of-the-art communication and networking technologies that have not been incorporated in 5G standards yet. The main reasons are related to both supply and demand. From the supply side, some technologies still require experimental verification and in-depth tests in practical environments. Meanwhile, the high cost and unsatisfactory backward compatibility also prevent them from being used. From the demand side, the services and devices supported by some advanced communication and networking technologies are not widely in demand.

Although 5G has adopted a gradual evolution strategy that is able to provide much more and better services than 4G, there is no ground-breaking technology in 5G. Instead, it inherits the fundamental performance enhancement mechanisms since 4G, and performance gains are achieved through an investment in more spectral and hardware resources  \cite{5741160}.

\subsection*{\textbf{Current Research Progress Towards 6G}}
A number of researchers have already provided visions for 6G and a series of advanced research planning activities have begun \cite{8412482,8603730,8631208,8869705,8792135,tariq2019speculative}. In the 6G vision and requirements suggested in \cite{8412482},  special attention is paid to the battery lifetime of mobile device and service classes in 6G, rather than data rate and latency.  In \cite{8603730}, it is pointed out that the communication system research in the post-5G era must incorporate with circuit and device manufacturing capabilities so as to form a closed feedback loop of research activities. A number of new communication scenarios in future networks around 2030 are predicted in \cite{8631208}, which encompass holographic calls, flying networks, teleoperated driving, and the tactile Internet. Further, it is foreseen that the same level of reliability as wired communications will be offered to future wireless communications.

 \cite{8869705} and \cite{8792135} summarize the future driving applications and trends as well as enabling technologies in 6G
networks. In particular, network decentralization based on blockchain technology is believed to be a key to simplify network management and provide satisfactory performance in 6G. The concept of human-centric service is also proposed and viewed as the emphasis in 6G. The key performance indicators (KPIs) of 6G are defined, and a speculative comparison between 5G and 6G is provided in \cite{tariq2019speculative}. 

Practical implementations, multiple access, air interface and data centre for 6G communications are envisioned and discussed in \cite{clazzer20195g}, \cite{miscopein2019air} and \cite{8751363}, respectively. Networking patterns of 6G networks are outlined in \cite{giordani2019towards, yanikomeroglu2018integrated,yaacoub2019key}, in which cell-less architecture, decentralised resource allocation, and three-dimensional super-connectivity are highly expected to exist in 6G networks. MTCs and vertical-specific wireless network solutions for 6G are studied in \cite{mahmood2019six}, which believes that 6G would facilitate the first wall-breaking standard to completely replace existing industry-specific communication standards and provide a unified solution enabling seamless connectivity for all needs in vertical industries.

Among all technological works pertaining to 6G, terahertz (THz) communications, artificial intelligence (AI), and reconfigurable intelligent surfaces are the most eye-catching protagonists. They are viewed as paradigm-shifting and revolutionary technologies in wireless communications. A comprehensive study of THz communications for 6G is reported in \cite{8732419}, which includes a detailed technological overview, transmitter-receiver designs, and various practical demonstrations. AI empowered 6G is believed to be able to provide a series of new features, e.g., self-aggregation, context-awareness, self-configuration, and opportunistic setup \cite{stoica20196g}. Additionally, AI empowered 6G would unlock the full potential of radio signals and enable the transformation from cognitive radio (CR) to intelligent radio (IR) \cite{8808168}. Machine learning is in particular crucial for realizing AI empowered 6G from the algorithmic perspective, which has been detailed in \cite{8681450}. Besides the algorithms, reconfigurable intelligent surfaces are supposed to be used to construct the hardware foundation of AI in wireless communications \cite{di2019smart}. Reconfigurable intelligent surfaces are also envisaged as the massive MIMO 2.0 in 6G and analysed in \cite{zhao2019survey,nadeem2019large,nadeem2019intelligent}. These attractive materials can also incorporate with index modulation (IM) to yield an increase in spectral efficiency in 6G networks \cite{basar2019large}.

Apart from above-released works, a number of 6G projects have already been started around the world, which aim to attain the initiative, define 6G, and reshape the framework as well as the business model of wireless communications. The first project refers to the 6Genesis Flagship Program (6GFP), a recently formed Finish consortium, which is followed by Terabit Bidirectional Multi-user Optical Wireless System (TOWS) for 6G LiFi started at the beginning of 2019. In March 2019, the first 6G Wireless Summit was held in Levi, Finland and formally triggered the starting gun of 6G research race in academia. Besides the summit, a number of small-scale workshops and seminars were also held worldwide to discuss the possibility of 6G, e.g., Huawei 6G Workshop, Wi-UAV Workshop of Globecom 2018, and Carleton 6G Workshop.

Except for academia, 6G and future networks also attract standardizing bodies, industrial organizations, and governments. IEEE launched IEEE Future Network with the tagline `Enabling 5G and Beyond' in August 2018. ITU-T Study Group 13 also established the ITU-T Focus Group Technologies for Network 2030 intending to understand the service requirements for future networks round 2030. Project Loon was triggered by Google and is now running independently, which plans to provide reliable Internet connection to the unconnected five billion population. A research group based on Terranova is now working toward the reliable 6G connection with 400 Gbit/s transmission capability in the THz spectrum. LG Electronics also announced the foundation of 6G Research Centre at Korea Advanced Institute of Science and Technology (KAIST), Daejeon, South Korea. Samsung kicked off its 6G research in June 2019. SK Telecom has decided to collaborate with Nokia and Ericsson in 6G research in the mid of 2019. In late 2018, China's Ministry of Industry and Information Technology declared the ambition of leading the wireless communication market around 2030 by expanding the research investment in 6G. Federal Communications Commission of the U.S. opened 95 GHz to 3 THz spectra for the use of 6G research, which marks the participation of the U.S., the world's biggest economic entity, in the 6G research race. In addition, an EU/Japan project under the ICT-09-2017 H2020 called `Networking Research beyond 5G' also investigates the possibility of using THz spectrum from 100 GHz to 450 GHz. With more details, we also summarize the country-wise research initiatives to achieve 6G in Table \ref{country}.

\begin{table*}[!t]
\renewcommand{\arraystretch}{1.3}
\caption{Country-wise research initiatives to achieve 6G.}
\label{country}
\centering
\begin{tabular}{c|c|c}
\hline
Country & Research initiative & Year\\
\hline\hline
Finland & \begin{tabular}{@{}c@{}} Finnish 6G research activity is coordinated by the University of Oulu, where an 6G initiative is launched \\ \end{tabular}
 & 2018\\
 \hline
USA & FCC opened the spectrum between 95 GHz and 3 THz to create a new category of experimental licenses. & 2019\\
\hline
S. Korea & \begin{tabular}{@{}c@{}} LG Electronics established a 6G research center in collaboration with KAIST. \\ ETRI has signed a memorandum of understanding with the University of Oulu in Finland to develop the 6G network technology.\\ Samsung Electronics Co. and SK Telecom Co. work together to develop technologies and business models related to 6G\\ SK Telecom Co. signed agreements with Finnish firm Nokia and Sweden's Ericsson to step up collaboration 6G network R\&D.\end{tabular}& 2019\\
\hline
China & \begin{tabular}{@{}c@{}}  The Ministry of Science and Technology planned to set up two working groups to carry out the 6G research activities:\\ The first group is from government departments to promote how 6G research and development will be carried out; \\ The second group is made from 37 universities, research institutes and companies, focusing on the technical side of 6G. \end{tabular}& 2019\\
\hline
Japan & \begin{tabular}{@{}c@{}} Japan readies \$ 2 billion to support industry research on 6G technology. \\ NTT and Intel have decided to form a partnership to work on 6G mobile network technology. \end{tabular}
 & TBD\\
\hline
\end{tabular}
\end{table*}

\section*{\textbf{Potential Application Scenarios and Challenges}}

6G communications are expected to provide improved services in terms of coverage, data rate and allow users to connect each other everywhere. It is expected to adopt unconventional communication networks to access several types of data and transmit them through conventional improved radio frequency (RF) networks, allowing new communication experience with virtual existence and involvement anywhere. To explicitly define the probable features of 6G communications, we foresee the potential application scenarios and challenges for 6G communications in this section. It should be noted that as a speculative study of 6G (`What Might 6G Be'), we intend to cover a large range of heated topics discussed in recently published works and conference releases, but with our own thoughts and comments to appraise these 6G candidate technologies.

\subsection*{\textbf{Enhanced Conventional Mobile Communications}}
As we declared at the beginning, 6G communications should be human centric, which implies that the conventional mobile communications will still hold the position of protagonist in 6G, in which classic cellular phone is the major tool of mobile communications. The challenges regarding conventional mobile communications comes from five aspects: 1) how to enhance security and protect privacy; 2) how to expand network coverage in a rapid and cost-efficient way, especially in distant and isolated areas; 3) how to reduce the cost of mobile communications; 4) how to extend the battery life of the mobile device; 5) how to provide a higher data rate with a lower end-to-end latency.

\subsection*{\textbf{Accurate Indoor Positioning}}
With the help of the Global Positioning System (GPS), outdoor positioning becomes full-fledged and can be regarded as accurate in most application scenarios now. However, indoor positioning is still far from maturity, because of the complex indoor electromagnetic propagation environment \cite{6837067}. Accurate and reliable indoor positioning services will radically change the living habits of mobile users and open up new niches for economic prosperity. On the other hand, there is a growing consensus that accurate indoor positioning might not be viable by sole utilizing RF communications \cite{dang2019enabling}. Such a crucial and impactful application is highly expected to be realised in the era of 6G with more advanced non-RF communication technologies.

\subsection*{\textbf{New Communications Terminals}}
In addition to the classic mobile phone and tablet, it is foreseen that there will be an increasing number of new communication devices in the 2030s. These new communication devices can be wearable devices, integrated headsets, and implantable sensors \cite{ullah2012comprehensive}. Different from the portable phone and tablet, these emerging devices impose diverse environmental and system requirements on communication networks. For example, there must be strict constraints on transmit power and frequency band used in these devices for health reasons. The device weight will become more sensitive when designing wearable devices and integrated headsets. Reliable power supply and security for implementable sensors are of high importance. In addition, there should be major dissimilitudes in mathematical modelling between these emerging communication devices and classic mobile phones and tablets.

\subsection*{\textbf{High-Quality Communication Services on Board}}
Despite the effort and endeavour of researchers in the 4G and 5G eras, it is undeniable that communication services on board are still unsatisfied in most cases nowadays. The communication services provided on board are challenging by the high mobility, Doppler shift, frequent hand-over, lack of coverage, and so on \cite{6403859}. Satellite communications enable communication services on board with acceptable service quality, but are too costly, especially in aircraft cabins \cite{8432390}. To provide high-quality communication services on board, not only new communication technologies must be employed in 6G communications, but also novel networking architectures shall be in use.

\subsection*{\textbf{Worldwide Connectivity and Integrated Networking}}
In the last decade, researchers drew attention to the communication services in dense metropolitan areas, especially for indoor communication scenarios. However, it should not be omitted that there is a large population around the world having no access to basic data services, especially in sparse, developing, and rural areas \cite{philbeck2017connecting}. The advent of the astonishing 6G era should not only benefit the majority in dense areas, but be shadowed to less dense areas. Making wireless networks not only vertical but also horizontal would benefit a much larger population. In this regard, worldwide connectivity is expected to be realised in 6G communications by a low-cost implementation scheme in order to guarantee the communication fairness of minority in sparse areas. Providing this service is greatly dependent on novel networking technologies.

To achieve the goal of worldwide connectivity, three-dimensional integrated networking would be utilised, which encompasses terrestrial, airborne, and satellite communications \cite{8473416}. Apart from satellite communications, most existing communication and networking architectures only consider two-dimensional scenarios, in which the heights of communication nodes are negligible \cite{8110602}. This modelling assumption is appropriate and efficient for 5G application scenarios. However,  it is envisioned that communications of flying nodes for achieving worldwide connectivity become ubiquitous in the 2030s and shall be taken into consideration when planning 6G networks. Such a three-dimensional integrated network  could bring considerable performance gains and unprecedented services to users \cite{8869705}. 

Apart from the communications on ground and over sky, extending the communication network to underwater environment is a crucial and even necessary element of worldwide connectivity, especially that more than $70 \%$ of the earth's surface is covered by water, and several marine applications needs live monitoring \cite{saeed2019underwater}. Underwater optical wireless communication (UOWC) can play a vital role in establishing reliable high data rate links with the help of acoustic communications \cite{7593257}. Underwater communication nodes such as autonomous vehicles, sensors and divers can be connected by underwater BSs using UOWC. Moreover, underwater communication networks are connected to terrestrial networks via water surface networks and aerial networks. Securing a sustainable energy source by wind, sun and water flow is an essential requirement for both underwater and water surface networks.

\subsection*{\textbf{Communications Supporting Vertical Industries}}
To deeply serve the physical world, 6G communications are also expected to continuously support the applications in vertical industries, including building and factory automation, manufacturing, e-health, transportation, agriculture, surveillance, and smart grid. These applications are essential to the Industry 4.0 and believed to be true paradigm shifts \cite{7980645}. They pose special service requirements in addition to conventional mobile telephony and broadband data communications. In particular, these vertical industries normally necessitate high standards with respect to connection reliability, transmission latency, and security \cite{whitepaper5gvertical}. To integrate these vertical industries in 6G communications, mMTC in 5G needs to be upgraded. More MTC application scenarios and types of machine nodes are required to be considered. 

In general, these vertically industrial applications can be classified into robotic communications and vehicular communications. Robotic communications are related to the communications of kinesthetic robotics and manufacturing robotics. Because any error, delay and malicious action in robotic communication applications could result in severe instability, robotic communications are reliability-critical, delay-critical and security-critical. Moreover, a huge number of heterogeneous data streams are generated in robotic communication networks, which yield a challenge for the current centralized networking architecture \cite{8294162}. 

For vehicular communications, two emerging technological trends in the vehicular industry are reshaping the physical world, which are corresponding to the self-driving and remote-driving technologies \cite{8450539}. Due to the development of both technologies in recent years, it is believed that they will be technically mature and widely applied before 2030. To enable both driving technologies in practice, massive vehicle-to-everything (V2X) communications must be studied and incorporated in 6G, which provide the basis for high-reliability and low-latency as well as secure exchange of massive driving and ambient data.

\subsection*{\textbf{Holographic Communications}}
6G is expected to be a conversion point from the traditional video conferencing to a virtual in-person meeting. To this end, a realistic projection of real-time movement needs to be transferred in negligible time, which resorts to holographic communications \cite{wakunami2016projection}. In fact, transferring three-dimensional image along the voice is not sufficient to convey the in-person presence. There is a need to have a three-dimensional video with stereo audio that can be reconfigured easily to capture several physical presences in the same area. In other words, one can interact with the received holographic data and modify the received video as needed. All this information needs to be captured and transmitted over reliable communication networks that should have an extremely large bandwidth.

\subsection*{\textbf{Tactile Communications}}
After using holographic communication to transfer a virtual vision of close-to-real sights of people, events, environments, and etc. It is beneficial to remotely exchange the physical interaction through the tactile Internet in real time  \cite{7403840}. Specifically, the expected services include teleoperation, cooperative automated driving, and interpersonal communication, where it should be possible to apply haptic control through communication networks. Efficient cross-layer communication system design has to be conducted to meet these stringent requirements of these applications. For example, new physical layer (PHY) schemes need to be developed, such as to improve the design of signalling systems, waveform multiplexing, and etc. As for the delay, all delay sources should be treated carefully, including buffering, queuing, scheduling, handover and the ones induced from protocols. Existing wireless communication systems cannot fulfil these requirements, and there is a necessity to over-the-air fibre communication systems  \cite{8474959}.

\subsection*{\textbf{Human Bond Communications}}
6G is expected to widely support the human-centric communication concept, where the human can access and/or share physical features. Human bond communication concept is proposed to allow accessing the human five senses \cite{prasad2016human}. Recently, the concept is expanded with the help of `communication through breath' scheme to allow reading the human bio-profile using the exhaled breath and even interact with the human body by inhalation using volatile organic compounds \cite{8647108}. As a result, it is possible to diagnose diseases, detect emotions, collect biological features and interact with the human body in a remote way. Developing communication systems that can replicate the human senses and human biological features requires interdisciplinary research. It is expected to have hybrid communication technologies that are able to sense different physical quantities and then share it with the intended receiver in a secure manner.

\subsection*{\textbf{Summary}}

\begin{figure}[!t]
\centering
\includegraphics[width=3.5in]{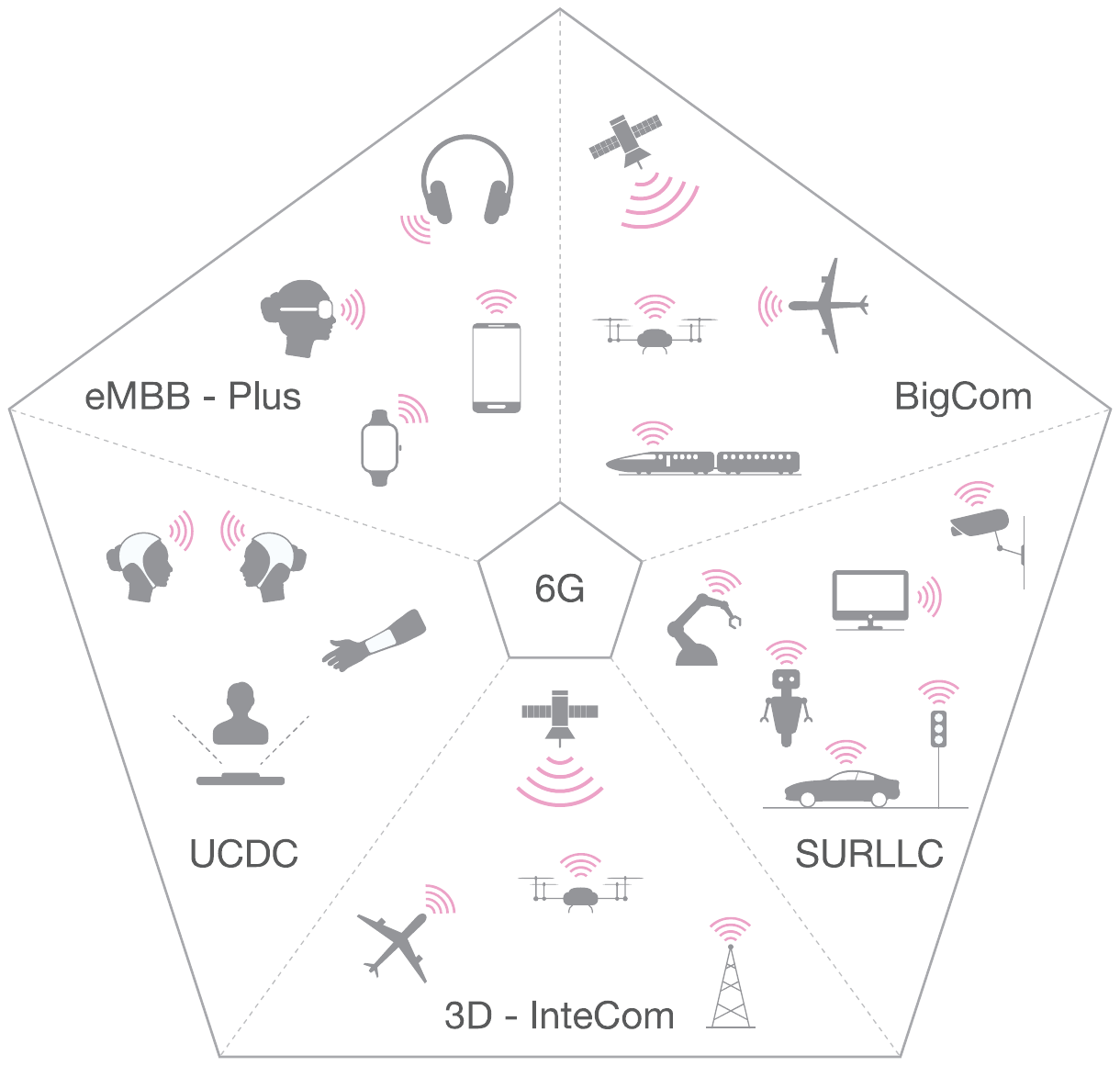}
\caption{\textbf{Five application scenarios supported by 6G communications.} The eMBB-Plus supports high-quality conventional mobile communications. The BigCom supports basic communications for remote areas. The SURLLC in 6G is a joint upgrade of the URLLC and the mMTC in 5G with higher requirements of reliability. The 3D-InteCom raises the network optimization and planning dimension to three. The UCDC provides the possibility to incorporate novel communication prototypes and paradigms.}
\label{6Gslices}
\end{figure}

Summarised from the above application scenarios, we anticipate five application scenarios supported by 6G communications: Enhanced Mobile Broadband Plus (eMBB-Plus), Big Communications (BigCom), Secure Ultra-Reliable Low-Latency Communications (SURLLC), Three-Dimensional Integrated Communications (3D-InteCom), Unconventional Data Communications (UCDC). These five application scenarios are pictorially illustrated in Fig. \ref{6Gslices}. We explain each of them in the following paragraphs.

The eMBB-Plus in 6G is the successor of the eMBB in 5G, serving the conventional mobile communications with much higher requirements and standards. It should also be more capable of optimizing the cellular networks in terms of interference, hand-over, as well as big data transmission and processing. Additional functionality will also be provided with an affordable expense to subscribers, e.g., accurate indoor positioning and globally compatible connection among diverse mobile operating networks. Most importantly, special attention of security, secrecy, and privacy shall be paid to the eMBB-Plus communication services.

Different from 5G that emphasises extremely good communication services in dense areas but to some extent neglects the service in remote areas, the BigCom in 6G cares about the service fairness between dense and remote areas. To be feasible, the BigCom does not intend to provide equally good services in both areas but keep a better resource balance. At least, the BigCom guarantees that the network coverage has to be large enough so as to provide acceptable data service wherever the communication subscribers are living or moving to. The Gini index and the Lorenz curve could be involved to evaluate the service fairness provided by the BigCom and should be treated as crucial indicators of UE in 6G \cite{6517050}.

The SURLLC in 6G is a joint upgrade of the URLLC and the mMTC in 5G, but with higher requirements of reliability (higher than 99.9999999\%, i.e., `Seven Sigma' from the viewpoints of quality control and process improvement) and latency (less than 0.1 ms) \cite{mahmood2019six}, as well as an additional demand on security. The SURLLC mainly serves the industrial and military communications, e.g., a variety of robots, high-precision machine tools, and conveyor systems in the 6G era. In addition, vehicular communications in 6G could also greatly benefit from the SURLLC.

The 3D-InteCom in 6G stresses that the network analysis, planning and optimization shall be raised from two dimensions to three dimensions, by which the heights of communications nodes must be taken into consideration. Satellite UAV, and underwater communications can be the examples of this three-dimensional scenario and benefit from three-dimensional analysis, planning and optimization. Accordingly, the analytical framework constructed for two-dimensional wireless communications stemmed from stochastic geometry and graph theory needs to be updated in the era of 6G \cite{5226957}. Considering the node height also enables the implementation of elevation beamforming with full-dimensional MIMO architectures, which provides another direction for network optimization \cite{nadeem2018elevation}.

The UCDC is probably the most open-ended application scenario in 6G communications. We intend to propose this application scenario to cover those novel communication prototypes and paradigms that cannot be classified into another four application scenarios. Currently, the definition and embodiment of the UCDC is still awaiting further exploration, but it should at least cover holographic, tactile, and human bond communications.

\section*{\textbf{Key Features and Enabling Communication Technologies of 6G}}
Based on the application scenarios and challenges as well as the five supported application scenarios in 6G discussed in the last section, we are now able to define the key features of 6G in this section. To enable the key features of 6G, multiple state-of-the-art communication technologies must be jointly applied, which are also summarised in this section. 

\begin{figure}[!t]
\centering
\includegraphics[width=2.8in]{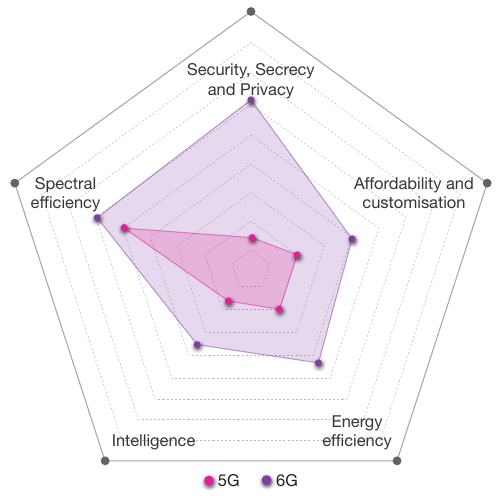}
\caption{\textbf{Qualitative comparison between 5G and 6G communications.} The comparison is made in terms of: security, secrecy, and privacy; spectral efficiency; intelligence; energy efficiency; and affordability and customisation. }
\label{radar}
\end{figure}

To begin with, a qualitative comparison between 5G and 6G communications is summarised in Fig. \ref{radar}.  In this figure, we first suppose that the spectral efficiency in 5G has already been close to the boundary by the advances in massive MIMO, network densification, and millimetre-wave transmission as well as a set of legacy multiplexing techniques inhering from 4G. As bounded by the Shannon limit, the spectral efficiency in 6G would hardly be improved on a large scale. In contrast, security, secrecy, and privacy in 6G communications should be significantly enhanced by new technologies. In 5G networks, traditional encryption algorithms based on the Rivest-Shamir-Adleman (RSA) public-key cryptosystems are still in use to provide transmission security and secrecy. The RSA cryptosystems have become insecure under the pressure of Dig Data and AI technologies, let alone privacy protection mechanisms that are far from being full-fledged in the 5G era. Incremental improvements would happen for energy efficiency, intelligence, affordability, and customisation. The energy efficiency gain would be accomplished by the maturity of energy harvesting technology and green communications. Intelligence in 6G can be classified into operational, environmental, and service levels, which will benefit from the thrust in AI developments. The improvements on affordability and customisation rely on novel networking architectures, promotion and operational strategies on the  market.  To be more specific, we give detailed comparisons from 1G to 6G communications in Table \ref{gcompare}.

\begin{table*}[!t]
\renewcommand{\arraystretch}{1.3}
\caption{Detailed comparisons from 1G to 6G communications.}
\label{gcompare}
\centering
\begin{tabular}{c|c|c|c|c|c|c}
\hline
Features & 1G & 2G & 3G & 4G & 5G & \begin{tabular}{@{}c@{}} 6G \\ (supposed) \end{tabular}\\
\hline\hline
Period & 1980-1990 & 1990-2000 & 2000-2010 & 2010-2020 & 2020-2030 & 2030-2040 \\
\hline
\begin{tabular}{@{}c@{}} Maximum\\rate \end{tabular} & 2.4 Kbps & 144 Kbps & 2 Mbps & 1 Gbps& 35.46 Gbps & 100 Gbps\\
\hline
\begin{tabular}{@{}c@{}} Maximum\\frequency \end{tabular} & 894 MHz& 1900 MHz & 2100 MHz & 6 GHz & 90 GHz & 10 THz\\
\hline
\begin{tabular}{@{}c@{}} Service\\level \end{tabular} & Voice & Text & Picture & Video & 3D VR/AR & Tactile\\
\hline
Standards & \begin{tabular}{@{}c@{}} MTS, AMPS,\\IMTS, PTT \end{tabular} & \begin{tabular}{@{}c@{}} GSM, IS-95,\\CDMA, EDGE \end{tabular} & \begin{tabular}{@{}c@{}} UMTS, WCDMA,\\IMT2000,\\ CDMA2000,\\ TD-SCDMA \end{tabular}& \begin{tabular}{@{}c@{}} WiMAX, LTE,\\ LTE-A \end{tabular} & \begin{tabular}{@{}c@{}} 5G NR,\\WWWW \end{tabular} & --\\
\hline
Multiplexing & FDMA & FDMA, TDMA & CDMA & OFDMA & OFDMA & \begin{tabular}{@{}c@{}} Smart OFDMA \\Plus IM \end{tabular} \\
\hline
Architecture & SISO & SISO & SISO & MIMO & \begin{tabular}{@{}c@{}} Massive\\MIMO \end{tabular} & \begin{tabular}{@{}c@{}} Intelligent\\ surface \end{tabular}  \\
\hline
\begin{tabular}{@{}c@{}} Core\\ network \end{tabular} & PSTN & PSTN & \begin{tabular}{@{}c@{}} Packet\\ N/W \end{tabular}& Internet& Internet, IoT& IoE \\
\hline
Highlight & Mobility & Digitization & Internet & \begin{tabular}{@{}c@{}} Real-time\\streaming \end{tabular} & Extremely high rate & \begin{tabular}{@{}c@{}}Security, secrecy, \\privacy \end{tabular}\\
\hline
\end{tabular}
\end{table*}


\subsection*{\textbf{High Security, Secrecy, and Privacy}}
Researchers placed great emphasis on network throughput, reliability, latency, and the number of served users in 4G and 5G communications. It has also been widely recognised that the two most efficient ways to improve these metrics are to densify the network and use a higher frequency to transmit signals \cite{5741160}. However, the security, secrecy, and privacy issues of wireless communications have been, to some extent, overlooked in the past decades. To protect data security, the classic encryption based on RSA algorithms is being challenged by increasingly powerful computers \cite{chen2016report}. PHY security technologies and quantum key distribution via visible light communications (VLC) would be the solutions to the data security challenge in 6G \cite{5751298,harrison2011quantum,8669813}. More advanced quantum computing and quantum communication technologies might also be deployed to provide intensive protection against various cyber attacks \cite{6576336}. Meanwhile, communication/data service providers have legally collected an enormous amount of user information, and private data leakage incidents happened occasionally. This becomes an unstable factor in the human-centric 6G network and could lead to a disastrous consequence without proper countermeasures. To solve this problem, it is envisioned that complete anonymization, decentralization, and untraceability can be realised in 6G networks by blockchain technology \cite{8425613}.

\subsection*{\textbf{High Affordability and Full Customisation}}
Again, from a human-centric perspective, technological success should not directly or indirectly increase the financial burden or deprive users' options. Therefore, high affordability and full customisation should be two important technological indicators of 6G communications.  The former is always ignored in most existing works. One might find their proposed solutions/schemes having a much higher transmission rate and/or reliability, but the cost rendered by such improvements will completely restrict their implementations in real life. The academic research activities for 6G should always try to get rid of such speciousness and endeavour to provide high affordability. 

Full customisation allows users to choose the service modes and adjust individual preference. For example, a user would like to have a low-rate but reliable data service; another user tolerates unreliable data service in order to get a lower communication expense  in return; some might only care about the energy consumption of their devices; some even intend to get rid of smart functionality due to the concerns of data security and privacy. All users will be granted the right to choose what they like in 6G, which should not be deprived by intelligent technologies and unnecessary system configurations. Accordingly, the performance analysis of 6G communication systems should also integrate multiple performance metrics into a whole, instead of treating them independently. UE would be explicitly defined and adopted as a pivotal metric for performance evaluation in the 6G era. That is, unlike 1G-5G, for which we added more elements in the quality of service (QoS) vector, we should map all required performance metrics as a whole to a unique UE performance metric for each individual user in 6G.

\subsection*{\textbf{Low Energy Consumption and Long Battery Life}}
The daily charging requirements of smartphones and tablets become annoying in 4G/LTE networks and will continue to the foreseeable 5G era. To release the daily charging constraint for most communication devices and facilitate communication services, low energy consumption and long battery life are two research emphases in 6G communications. To lower energy consumption, the computing tasks of a user device can be off-loaded to smart base stations (BSs) with reliable power supply or pervasive smart radio space \cite{8368232}. Cooperative relay communications and network densification would also help to reduce the transmit power of mobile device by reducing the per-hop signal propagation distance \cite{4600214,7010521}. To achieve a long battery life, various energy harvesting methodologies would be applied in 6G, which not only harvest energy from ambient radios, but also the energy from micro-vibrations and sunlight \cite{7010878}. Long-distance wireless power charging would also be a promising approach to extend battery life, but in-depth investigations are indispensable to avoid health related issues \cite{6757202}.

\subsection*{\textbf{High Intelligence}}
The high intelligence in 6G will benefit network operations, wireless propagation environments, and communication services, which refer to  operational intelligence, environmental intelligence, and service intelligence, respectively.
\subsubsection{Operational intelligence}
Conventional network operation involves a great number of multi-objective performance optimization problems subject to a series of complex constraints. Resources, including communication devices, frequency bands, transmit power, and so on are required to be arranged in a proper way so as to achieve a satisfactory level of network operation. Moreover, these multi-objective performance optimization problems are usually NP-hard, and optimal solutions are hard to be obtained on a real-time basis. With the development of machine learning techniques, especially deep learning, a BS equipped with graphics processing units (GPUs) or the control centre of core network could carry out relevant learning algorithms to allocate resources efficiently to achieve performance close to the optimum \cite{8382166}.

\subsubsection{Environmental intelligence}
Meanwhile, by the advances on smart radio space and smart materials, distributed and pervasive intelligence of the holistic communication environment, including wireless channels, would become possible \cite{4078950}. This could provide self-organizing and self-healing properties for the 6G network and enable reliable device-to-device (D2D) communications in a fully intelligent way. Latest works have defined and justified the conception of reconfigurable intelligent surfaces, which are designed to sense the wireless environment and apply customised transformations to the radio waves in an adaptive manner \cite{di2019smart,basar2019wireless}. This conception solidifies the hardware foundation of environmental intelligence.

\subsubsection{Service intelligence}
Furthermore, as a human-centric network, the high intelligence of 6G network also directly reflects in a plethora of communication services, e.g., indoor/outdoor positioning, multi-device management, information search, e-health, surveillance, cyber security \cite{8493126,dang2019enabling}. Service intelligence enables these services to be provided in a satisfactory and personalised way. For example, the accuracy of indoor positioning can be greatly improved by deep learning techniques \cite{8612930}, and personalised healthcare is realised by intelligent IoT and multi-model data collecting infrastructure \cite{7156004}. The service intelligence mainly benefit from high-performance core networks implemented in 6G \cite{8255764,kishk2019capacity}.

\subsection*{\textbf{Extremely Larger Bandwidth than 5G}}
The THz band defined from 0.1 THz to 10 THz was known as a gap band between the microwave and optical spectra \cite{8901159}. Nowadays, electronic, photonic and hybrid electronic-photonic based methods are developed \cite{sengupta2018terahertz}. Thus, hybrid THz/free space optical (FSO) systems are highly expected to be realised in 6G using the hybrid electronic-photonic transceivers, where an optical laser can be used to generate THz signal or send an optical signal. The hybrid link offers plenty of bandwidth for signal transmission and has the immunity to adverse weather conditions  \cite{nagatsuma2016advances}. THz transmission can play a vital role in the uplink, because a line-of-sight link is not required. Thus, THz uplink solution offers a reliable communication link for VLC networks compared with the infrared solution that needs tracking and positioning system. Hybrid VLC/THz system introduces robust communication solutions against ambient light that reduces the signal-to-noise ratio of the VLC system.

\subsection*{\textbf{Trade-offs Between Key Features and Potential Solutions}}
It should be noted that as an engineering system, it is not possible to satisfy all wished features without investing more resources, because there exist a number of trade-offs between these features. For 6G communications, we must figure out a way to invest adequate resources in guaranteeing some critical features and raise up all features with an equilibrium. To this end, we discuss a set of crucial trade-offs in 6G communications regarding these key features and the potential solutions in the following paragraphs.
\subsubsection{Privacy vs. intelligence}
As a human-centric network, the trade-off between privacy and intelligence would be the most important one in 6G communications. On the one hand, AI algorithms need to get access to personal data and process them, so as to optimise network operations, adapt network coefficients, and provide high-quality services. On the other hand, privacy would be sacrificed for the sake of high intelligence. A potential solution  is to introduce an intermediate agent between the end-user data and AI algorithms. Such an intermediate agent should be third-party and unmanned if possible, and operate on a decentralized basis. All private and sensitive data will be anonymised by this third-party agent and become untraceable in any way.

\subsubsection{Affordability vs. intelligence}
High intelligence introduces a high degree of system complexity, which could raise up the costs to network operators and device manufacturers. All these raised costs will finally be transferred to less affordable products to end users. To resolve this trade-off, technological breakthroughs in intelligent systems are necessary, but more importantly, a new commercial strategy would be helpful.  Once security, secrecy, and privacy are guaranteed, end users have the right to exchange the accessibility of their anonymised data for a lower data price. A similar feature of the smart grid, by which electricity users can also sell self-generated electricity back to electricity companies would be borrowed to 6G communication networks.

\subsubsection{Customisation vs. intelligence}
High intelligence provided by AI algorithms and smart devices weakens the free will of human beings. In other words, the user preference might not be always aligned with the optimised option produced by AI algorithms. The contradictory situation becomes severer when multiple users are taken into account. This conflict can be formulated as the trade-off between customisation and intelligence in 6G communications. In our opinion, the priority shall always be given to the customisation, and prohibitive clauses are on demand for AI algorithms and smart devices. These prohibitive clauses should be stipulated in the most fundamental and underlying protocols of 6G communications. In this way, intelligent services can only be provided within the permissible boundary. 

\subsubsection{Security vs. spectral efficiency}
Conventionally, to ensure a secure transmission, more spectral resource shall be in use for preventive measures, and the net load for transmitting information is lowered accordingly, given a limited radio spectrum. We have to recognise that this trade-off between security and spectral efficiency is difficult to resolve, but we can mitigate it in three possible ways. First, researchers might try to design a more efficient encryption algorithm. However, this direction would be rather difficult due to the maturity of data encryption. Second, researchers might resort to PHY security technologies for providing security protection without a great loss of spectral efficiency. Third, AI algorithms can also help to detect network anomaly and would be utilised in 6G networks to provide an early warning mechanism for security enhancement.

\subsubsection{Spectral efficiency vs. energy efficiency}
The trade-off between spectral efficiency and energy efficiency is a frequent topic in the field of wireless communications. The discussion pertaining to this well-known trade-off ran through all wireless generations and will, of course, be one of the focuses in 6G communications. However, different from 1G-5G, a new path-breaking technology would be introduced to greatly alleviate this trade-off, which is energy harvesting. By energy harvesting, user devices are capable of harvesting radio, vibratory, and solar energy from the ambient environment and the constraint on energy consumption can thereby be released. The environmental intelligence realised by ubiquitous intelligent surfaces would also help to mitigate the spectrum-energy trade-off by adapting radio propagation environments.

\section*{\textbf{Beyond the Communication Technologies}}
Communication technologies are crucial, but not all. To promote a new technological paradigm and make it socio-economically profitable, we must always keep the issues beyond technology in mind. In this section, we briefly discuss several crucial issues vis-\`{a}-vis 6G beyond the communication technologies \textit{per se}.

\subsection*{\textbf{Dependency on Basic Sciences}}
It is admitted that the advancement of wireless communications is highly restricted by basic sciences, especially mathematics and physics. As detailed in \cite{5741160}, we are squeezing the last lemon juice of Shannon's treatise published in 1948 and almost reach the hard wall set by information theory. What is worse, incapable mathematical tools prevent us from exploring the exact performance of communication systems and make us lost in the \textit{asymptopia}. As a result, a large number of impractical assumptions are made in order to make analysis mathematically tractable, which cannot provide much insight and guideline for 6G communications. The breakthrough in mathematics would often result in a new research boom in wireless communications, and one example is the stochastic geometry and graph theory applied for wireless network modelling \cite{5226957}. To summarize, researchers shall pay sufficient attention to basic sciences and interdisciplinary fields in order to realise 6G networks.

\subsection*{\textbf{Dependency on Upstream Industries}}
In the wireless communication research community, it is widely agreed that the most efficient ways to enhance wireless communication systems are to expand the usage to high-frequency spectrum and to reduce the coverage of a single cell \cite{5741160}. The former tendency is witnessed by the evolution from cellular radio spectrum to millimetre-wave spectrum, THz spectrum, and visible light spectrum. The later tendency refers to the network densification. On the other hand, both tendencies must match up the developments in upstream industries, e.g., electronics manufacturing. First of all, in theoretical research, one can assume an arbitrarily high frequency for use, but in reality, the communication devices constituted realistic electronic components must be able to meet these requirements. In some cases, the resultant data rate has even exceeded the allowable constraint on the electronic circuit by the current manufacturing level, or the signal on higher frequency bands cannot be proceeded by currently commercial chips at all. Neglecting the dependency on upstream industries will turn the 6G research to be nothing but a theoretical carnival.

\subsection*{\textbf{Demand-Oriented Research Roadmap}}
It has been noticed that there exists a visible mismatch of PHY research activities in industry and academia \cite{4468730}. As suggested in \cite{8603730} and \cite{5741160}, a closer connection between industrial and academic researches should be constructed so as to form a positively closed feedback loop for adjusting research roadmap. More directly, such a positively closed feedback loop can be extended to the market and the end beneficiaries of 6G. In this way, a demand-oriented research roadmap can be well designed and adapted in a much more effective and efficient manner. To achieve this goal, it is required to introduce the ideas of \textit{value engineering} to plan academic research activities. In this way, 6G research roadmap should not be defined by the technological embodiments, but by the function and cost as a whole from a value engineering view. In other words, the research activities in 6G should not simply aim at adding more functions without considering the cost and demand from the end beneficiaries' perspective, but target the \textit{value} of the implemented service. Specifically, end beneficiaries shall be granted the right to have their  voices to reshape the research roadmap in the 6G era. To well satisfy the  demands of multiple stakeholders and bridge between academic and real-world problems, the barrier among various disciplines should be removed. More economic and sociological methodologies, e.g, empirical analysis and PESTEL analysis (PESTEL: an acronym that stands for \textbf{P}olitical, \textbf{E}conomic, \textbf{S}ocial, \textbf{T}echnological, \textbf{E}nvironmental and \textbf{L}egal factors), could be introduced for appraising and tailoring 6G research roadmap.

\subsection*{\textbf{Business Model and Commercialization of 6G}}
Previous research activities primarily focus on the technology itself, but rarely on the business model and commercialization. Omitting the marketing aspects would lead to failure (3G could be to some extent an example of such failure \cite{8412482}). Network densification is a promising solution to satisfy the data transmission burst, but who should pay for it? Building new BSs is costly after all, because of land use right granting and construction operations. Moreover, as 6G communications would bring ground-breaking communication technologies relying on novel network architectures, how to ensure the backward compatibility of 6G with 4G/LTE, Wi-Fi and 5G is still questionable and worth investigating. The overall cost for updating the existing infrastructures for 6G communications needs to be evaluated first, and then the business model and commercialization of 6G should be studied for its commercial triumph. One should always remember that for most ordinary users and government policy makers, paying several times higher to get a dispensable performance gain in terms of transmission rate or latency will not be accepted, let alone appreciated.

\subsection*{\textbf{Potential Health and Psychological Issues for Users}}
The `base station myth' is a frequent topic in public media and could even trigger severe protests \cite{drake2006mobile}, which reflects the health concerns of users about radiation safety. As reported, electromotive force (EMF) limits have been reached in some cities. With a densified network with a smaller coverage per BS and the use of higher frequency band, there are reasons to believe that such concerns will be aggravated in the era of 6G. As 6G communications are expected to be human centric, special attention must be paid to the potential health issues brought to users. In this context, EMF-aware transmission would be a novel concept to be introduced in 6G to mitigate the health concerns \cite{6807642}. Bandwidth expansion from the millimetre-wave regime to the THz regime also causes uncertain biological impacts on humans and animals. Careful studies are required to examine the safety  of THz radiation \cite{8732419}.

Apart from health issues, the psychological barrier would also be a factor hindering the large-scale implementation of 6G from a human-centric perspective. As envisioned in some proposals even for 5G networks, massive sensors are deployed all over the space, and they are smart to detect, understand, communicate, and respond (fortunately, such a sci-fi scene has been greatly exaggerated). Then, the question will be: will people really enjoy and be comfortable to live in such a smart space? Will people be delighted to be recorded and watched by such a technocratic `big brother'? Without a careful study on these psychological issues before implementing in practice, 6G could cause catastrophic consequences and even deconstruct existing trust in ICTs \cite{pieters2011explanation}. 6G is expected to be not only technologically trustworthy \cite{8760250}, but also societally trustworthy.

\subsection*{\textbf{Social Factors Hindering the Worldwide Connectivity}}
As pointed out in the background paper of the World Economic Forum at Davos Annual Meeting 2017 \cite{philbeck2017connecting}, apart from technological and economic factors, social factors could also prevent worldwide connectivity in 6G. That is, the people living in developing areas are not motivated to be connected, because of the lack of content relevance, language barrier, and computer literacy. This is mainly a demand-side issue and shall be given full consideration when deploying 6G networks for worldwide connectivity. Incentive schemes and campaigns sponsored by local governments and private companies would be beneficial to persuade the unconnected in distant areas to be connected and promote the concept of worldwide connectivity in the 6G era. The promotions of e-payment and e-taxi in China are good examples that connect most people who never use smart phone before.

\section*{\textbf{Conclusions}}
We have provided a vision for 6G communications from a human-centric perspective that could serve as research guide in the post-5G era. We suggest that high security, secrecy, and privacy should be the key features of 6G, and should be given particular attention by the wireless research community. We envisioned and explained potential application scenarios that should be supported in 6G. We also introduced key features and enabling technologies for 6G communications. Finally, we discussed other crucial issues beyond communication technologies that should be considered in the development of 6G.

\section*{\textbf{Acknowledgement}}
Fig. \ref{evolution}, Fig. \ref{6Gslices}, and Fig. \ref{radar} were created by Ivan Gromicho, Scientific Illustrator at King Abdullah University of Science and Technology (KAUST).

\bibliographystyle{IEEEtran}

\begin{thebibliography}{10}
\providecommand{\url}[1]{#1}
\csname url@samestyle\endcsname
\providecommand{\newblock}{\relax}
\providecommand{\bibinfo}[2]{#2}
\providecommand{\BIBentrySTDinterwordspacing}{\spaceskip=0pt\relax}
\providecommand{\BIBentryALTinterwordstretchfactor}{4}
\providecommand{\BIBentryALTinterwordspacing}{\spaceskip=\fontdimen2\font plus
\BIBentryALTinterwordstretchfactor\fontdimen3\font minus
  \fontdimen4\font\relax}
\providecommand{\BIBforeignlanguage}[2]{{%
\expandafter\ifx\csname l@#1\endcsname\relax
\typeout{** WARNING: IEEEtran.bst: No hyphenation pattern has been}%
\typeout{** loaded for the language `#1'. Using the pattern for}%
\typeout{** the default language instead.}%
\else
\language=\csname l@#1\endcsname
\fi
#2}}
\providecommand{\BIBdecl}{\relax}
\BIBdecl

\bibitem{alsharif2017evolution}
M.~H. Alsharif and R.~Nordin, ``Evolution towards fifth generation ({5G})
  wireless networks: Current trends and challenges in the deployment of
  millimetre wave, massive mimo, and small cells,'' \emph{Telecommunication
  Systems}, vol.~64, no.~4, pp. 617--637, 2017.

\bibitem{8412482}
K.~{David} and H.~{Berndt}, ``{6G} vision and requirements: {Is} there any need
  for beyond {5G}?'' \emph{IEEE Vehicular Technology Magazine}, vol.~13, no.~3,
  pp. 72--80, Sept. 2018, \\ \textbf{ This is the first publication visioning
  6G from the perspective of service}.

\bibitem{8603730}
V.~{Raghavan} and J.~{Li}, ``Evolution of physical-layer communications
  research in the post-{5G} era,'' \emph{IEEE Access}, vol.~7, pp.
  10\,392--10\,401, 2019, \\ \textbf{This paper points out the potential
  research directions of physical-layer communications in the post-5G era}.

\bibitem{8631208}
A.~{Yastrebova}, R.~{Kirichek}, Y.~{Koucheryavy}, A.~{Borodin}, and
  A.~{Koucheryavy}, ``Future networks 2030: {Architecture} \& requirements,''
  in \emph{Proc. IEEE ICUMT}, Moscow, Russia, Nov. 2018, pp. 1--8, \\
  \textbf{This paper details the project of Future Networks 2030}.

\bibitem{8869705}
W.~{Saad}, M.~{Bennis}, and M.~{Chen}, ``A vision of {6G} wireless systems:
  Applications, trends, technologies, and open research problems,'' \emph{IEEE
  Network}, pp. 1--9, 2019.

\bibitem{8792135}
E.~{Calvanese Strinati} \emph{et~al.}, ``{6G}: {The} next frontier: From
  holographic messaging to artificial intelligence using subterahertz and
  visible light communication,'' \emph{IEEE Vehicular Technology Magazine},
  vol.~14, no.~3, pp. 42--50, Sept. 2019.

\bibitem{tariq2019speculative}
F.~Tariq \emph{et~al.}, ``A speculative study on {6G},'' \emph{arXiv preprint
  arXiv:1902.06700}, 2019.

\bibitem{7000985}
S.~{Chen}, J.~{Zhao}, and Y.~{Peng}, ``The development of {TD-SCDMA 3G} to
  {TD-LTE}-advanced {4G} from 1998 to 2013,'' \emph{IEEE Wireless
  Communications}, vol.~21, no.~6, pp. 167--176, Dec. 2014.

\bibitem{6769024}
J.~{Rissen} and R.~{Soni}, ``The evolution to {4G} systems,'' \emph{Bell Labs
  Technical Journal}, vol.~13, no.~4, pp. 1--5, 2009.

\bibitem{923566}
Y.~{Raivio}, ``{4G}-hype or reality,'' in \emph{Proc. International Conference
  on {3G} Mobile Communication Technologies}, London, UK, Mar. 2001, pp.
  346--350.

\bibitem{4468730}
M.~{Dohler}, D.~{Meddour}, S.~{Senouci}, and A.~{Saadani}, ``Cooperation in
  {4G} - hype or ripe?'' \emph{IEEE Tech. and Soc. Mag.}, vol.~27, no.~1, pp.
  13--17, Spring 2008.

\bibitem{Frias2012}
Z.~Frias and J.~P{\'e}rez, ``Techno-economic analysis of femtocell deployment
  in long-term evolution networks,'' \emph{EURASIP Journal on Wireless
  Communications and Networking}, vol. 2012, no.~1, p. 288, Sept 2012.

\bibitem{moral2011technoeconomic}
A.~Moral \emph{et~al.}, ``Technoeconomic evaluation of cooperative relaying
  transmission techniques in {OFDM} cellular networks,'' \emph{EURASIP Journal
  on Advances in Signal Processing}, vol. 2011, p.~6, 2011.

\bibitem{8750780}
Z.~{Wang}, S.~{Dang}, S.~{Shaham}, Z.~{Zhang}, and Z.~{Lv}, ``Basic research
  methodology in wireless communications: The first course for research-based
  graduate students,'' \emph{IEEE Access}, vol.~7, pp. 86\,678--86\,696, 2019.

\bibitem{6824752}
J.~G. {Andrews} \emph{et~al.}, ``What will {5G} be?'' \emph{IEEE Journal on
  Selected Areas in Communications}, vol.~32, no.~6, pp. 1065--1082, June 2014.

\bibitem{8258595}
S.~{Parkvall}, E.~{Dahlman}, A.~{Furuskar}, and M.~{Frenne}, ``{NR}: The new
  {5G} radio access technology,'' \emph{IEEE Communications Standards
  Magazine}, vol.~1, no.~4, pp. 24--30, Dec. 2017.

\bibitem{8638891}
M.~{Patzold}, ``{5G} is coming around the corner,'' \emph{IEEE Vehicular
  Technology Magazine}, vol.~14, no.~1, pp. 4--10, Mar. 2019, \\ \textbf{This
  editorial summarizes the latest achievements of 5G research deployment}.

\bibitem{5741160}
M.~{Dohler}, R.~W. {Heath}, A.~{Lozano}, C.~B. {Papadias}, and R.~A.
  {Valenzuela}, ``Is the {PHY} layer dead?'' \emph{IEEE Communications
  Magazine}, vol.~49, no.~4, pp. 159--165, Apr. 2011, \\ \textbf{This paper
  describers a number of commom issues that have lasted for a long time in the
  research community of wireless communications.}

\bibitem{clazzer20195g}
F.~Clazzer \emph{et~al.}, ``From {5G} to {6G}: Has the time for modern random
  access come?'' \emph{arXiv preprint arXiv:1903.03063}, 2019.

\bibitem{miscopein2019air}
B.~Miscopein, J.-B. Dor{\'e}, E.~Strinati, D.~Kt{\'e}nas, and S.~Barbarossa,
  ``Air interface challenges and solutions for future {6G} networks,'' 2019.

\bibitem{8751363}
S.~{Rommel}, T.~R. {Raddo}, and I.~T. {Monroy}, ``Data center connectivity by
  {6G} wireless systems,'' in \emph{Proc. IEEE PSC}, Limassol, Cyprus, Sept.
  2018, pp. 1--3.

\bibitem{giordani2019towards}
M.~Giordani, M.~Polese, M.~Mezzavilla, S.~Rangan, and M.~Zorzi, ``Towards {6G}
  networks: Use cases and technologies,'' \emph{arXiv preprint
  arXiv:1903.12216}, 2019.

\bibitem{yanikomeroglu2018integrated}
H.~Yanikomeroglu, ``Integrated terrestrial/non-terrestrial {6G} networks for
  ubiquitous {3D} super-connectivity,'' in \emph{Proceedings of the 21st ACM
  International Conference on Modeling, Analysis and Simulation of Wireless and
  Mobile Systems}.\hskip 1em plus 0.5em minus 0.4em\relax ACM, 2018, pp. 3--4.

\bibitem{yaacoub2019key}
E.~Yaacoub and M.-S. Alouini, ``A key {6G} challenge and
  opportunity--connecting the remaining 4 billions: A survey on rural
  connectivity,'' \emph{arXiv preprint arXiv:1906.11541}, 2019.

\bibitem{mahmood2019six}
N.~H. Mahmood \emph{et~al.}, ``Six key enablers for machine type communication
  in {6G},'' \emph{arXiv preprint arXiv:1903.05406}, 2019.

\bibitem{8732419}
T.~S. {Rappaport} \emph{et~al.}, ``Wireless communications and applications
  above 100 {GHz}: Opportunities and challenges for {6G} and beyond,''
  \emph{IEEE Access}, vol.~7, pp. 78\,729--78\,757, 2019.

\bibitem{stoica20196g}
R.-A. Stoica and G.~T.~F. de~Abreu, ``{6G}: the wireless communications network
  for collaborative and {AI} applications,'' \emph{arXiv preprint
  arXiv:1904.03413}, 2019.

\bibitem{8808168}
K.~B. {Letaief}, W.~{Chen}, Y.~{Shi}, J.~{Zhang}, and Y.~A. {Zhang}, ``The
  roadmap to {6G}: {AI} empowered wireless networks,'' \emph{IEEE
  Communications Magazine}, vol.~57, no.~8, pp. 84--90, Aug. 2019.

\bibitem{8681450}
S.~J. {Nawaz}, S.~K. {Sharma}, S.~{Wyne}, M.~N. {Patwary}, and
  M.~{Asaduzzaman}, ``Quantum machine learning for {6G} communication networks:
  State-of-the-art and vision for the future,'' \emph{IEEE Access}, vol.~7, pp.
  46\,317--46\,350, 2019.

\bibitem{di2019smart}
D.~Renzo \emph{et~al.}, ``Smart radio environments empowered by reconfigurable
  {AI} meta-surfaces: an idea whose time has come,'' \emph{EURASIP Journal on
  Wireless Communications and Networking}, vol. 2019, no.~1, p. 129, 2019.

\bibitem{zhao2019survey}
J.~Zhao, ``A survey of reconfigurable intelligent surfaces: Towards {6G}
  wireless communication networks with massive {MIMO} 2.0,'' \emph{arXiv
  preprint arXiv:1907.04789}, 2019.

\bibitem{nadeem2019large}
Q.-U.-A. Nadeem, A.~Kammoun, A.~Chaaban, M.~Debbah, and M.-S. Alouini, ``Large
  intelligent surface assisted {MIMO} communications,'' \emph{arXiv preprint
  arXiv:1903.08127}, 2019.

\bibitem{nadeem2019intelligent}
------, ``Intelligent reflecting surface assisted multi-user {MISO}
  communication,'' \emph{arXiv preprint arXiv:1906.02360}, 2019.

\bibitem{basar2019large}
E.~Basar, ``Large intelligent surface-based index modulation: A new beyond
  {MIMO} paradigm for {6G},'' \emph{arXiv preprint arXiv:1904.06704}, 2019.

\bibitem{6837067}
J.~{Oh}, M.~{Thiel}, and K.~{Sarabandi}, ``Wave-propagation management in
  indoor environments using micro-radio-repeater systems,'' \emph{IEEE Antennas
  and Propagation Magazine}, vol.~56, no.~2, pp. 76--88, Apr. 2014.

\bibitem{dang2019enabling}
S.~Dang, G.~Ma, B.~Shihada, and M.-S. Alouini, ``Enabling smart buildings by
  indoor visible light communications and machine learning,'' \emph{arXiv
  preprint arXiv:1904.07959}, 2019.

\bibitem{ullah2012comprehensive}
S.~Ullah \emph{et~al.}, ``A comprehensive survey of wireless body area
  networks,'' \emph{Journal of medical systems}, vol.~36, no.~3, pp.
  1065--1094, 2012.

\bibitem{6403859}
X.~{Li}, S.~{Hong}, V.~D. {Chakravarthy}, M.~{Temple}, and Z.~{Wu},
  ``Intercarrier interference immune single carrier {OFDM} via magnitude-keyed
  modulation for high speed aerial vehicle communication,'' \emph{IEEE
  Transactions on Communications}, vol.~61, no.~2, pp. 658--668, Feb. 2013.

\bibitem{8432390}
X.~{Zhang}, W.~{Cheng}, and H.~{Zhang}, ``Heterogeneous statistical {QoS}
  provisioning over airborne mobile wireless networks,'' \emph{IEEE Journal on
  Selected Areas in Communications}, vol.~36, no.~9, pp. 2139--2152, Sept.
  2018.

\bibitem{philbeck2017connecting}
I.~Philbeck, ``Connecting the unconnected: Working together to achieve connect
  2020 agenda targets,'' in \emph{Special session of the Broadband Commission
  and the World Economic Forum at Davos Annual Meeting}, 2017.

\bibitem{8473416}
R.~{Gopal} and N.~{BenAmmar}, ``Framework for unifying {5G} and next generation
  satellite communications,'' \emph{IEEE Network}, vol.~32, no.~5, pp. 16--24,
  Sept. 2018.

\bibitem{8110602}
S.~{Dang}, J.~P. {Coon}, and G.~{Chen}, ``Outage performance of two-hop {OFDM}
  systems with spatially random decode-and-forward relays,'' \emph{IEEE
  Access}, vol.~5, pp. 27\,514--27\,524, 2017.

\bibitem{saeed2019underwater}
N.~Saeed, A.~Celik, T.~Y. Al-Naffouri, and M.-S. Alouini, ``Underwater optical
  wireless communications, networking, and localization: A survey,'' \emph{Ad
  Hoc Networks}, p. 101935, 2019.

\bibitem{7593257}
Z.~{Zeng}, S.~{Fu}, H.~{Zhang}, Y.~{Dong}, and J.~{Cheng}, ``A survey of
  underwater optical wireless communications,'' \emph{IEEE Communications
  Surveys Tutorials}, vol.~19, no.~1, pp. 204--238, Firstquarter 2017.

\bibitem{7980645}
M.~{Dohler} \emph{et~al.}, ``Internet of skills, where robotics meets {AI},
  {5G} and the {Tactile Internet},'' in \emph{Proc. IEEE EuCNC}, Oulu, Finland,
  June 2017, pp. 1--5.

\bibitem{whitepaper5gvertical}
``{5G} communications for automation in vertical domains,'' 5G Americas, Tech.
  Rep., Nov. 2018.

\bibitem{8294162}
F.~{Voigtlander} \emph{et~al.}, ``{5G} for robotics: Ultra-low latency control
  of distributed robotic systems,'' in \emph{Proc. IEEE ISCSIC}, Oct. 2017, pp.
  69--72.

\bibitem{8450539}
N.~{Cheng} \emph{et~al.}, ``Big data driven vehicular networks,'' \emph{IEEE
  Network}, vol.~32, no.~6, pp. 160--167, Nov. 2018.

\bibitem{wakunami2016projection}
K.~Wakunami \emph{et~al.}, ``Projection-type see-through holographic
  three-dimensional display,'' \emph{Nature communications}, vol.~7, p. 12954,
  2016.

\bibitem{7403840}
M.~{Simsek}, A.~{Aijaz}, M.~{Dohler}, J.~{Sachs}, and G.~{Fettweis},
  ``{5G}-enabled tactile internet,'' \emph{IEEE Journal on Selected Areas in
  Communications}, vol.~34, no.~3, pp. 460--473, Mar. 2016.

\bibitem{8474959}
K.~S. {Kim} \emph{et~al.}, ``Ultrareliable and low-latency communication
  techniques for tactile {Internet} services,'' \emph{Proceedings of the IEEE},
  vol. 107, no.~2, pp. 376--393, Feb. 2019.

\bibitem{prasad2016human}
R.~Prasad, ``Human bond communication,'' \emph{Wireless Personal
  Communications}, vol.~87, no.~3, pp. 619--627, 2016.

\bibitem{8647108}
M.~{Khalid}, O.~{Amin}, S.~{Ahmed}, B.~{Shihada}, and M.-S. Alouini,
  ``Communication through breath: Aerosol transmission,'' \emph{IEEE
  Communications Magazine}, vol.~57, no.~2, pp. 33--39, Feb. 2019.

\bibitem{6517050}
H.~Shi, R.~V. {Prasad}, E.~{Onur}, and I.~G. M.~M. {Niemegeers}, ``Fairness in
  wireless networks: Issues, measures and challenges,'' \emph{IEEE
  Communications Surveys Tutorials}, vol.~16, no.~1, pp. 5--24, First 2014.

\bibitem{5226957}
M.~{Haenggi}, J.~G. {Andrews}, F.~{Baccelli}, O.~{Dousse}, and
  M.~{Franceschetti}, ``Stochastic geometry and random graphs for the analysis
  and design of wireless networks,'' \emph{IEEE Journal on Selected Areas in
  Communications}, vol.~27, no.~7, pp. 1029--1046, Sept. 2009.

\bibitem{nadeem2018elevation}
Q.-U.-A. Nadeem, A.~Kammoun, and M.-S. Alouini, ``Elevation beamforming with
  full dimension {MIMO} architectures in {5G} systems: A tutorial,'' \emph{IEEE
  Communications Surveys Tutorials}, 2019.

\bibitem{chen2016report}
L.~Chen \emph{et~al.}, \emph{Report on post-quantum cryptography}.\hskip 1em
  plus 0.5em minus 0.4em\relax US Department of Commerce, National Institute of
  Standards and Technology, 2016.

\bibitem{5751298}
Y.~{Shiu}, S.~Y. {Chang}, H.~{Wu}, S.~C.~. {Huang}, and H.~{Chen}, ``Physical
  layer security in wireless networks: a tutorial,'' \emph{IEEE Wireless
  Communications}, vol.~18, no.~2, pp. 66--74, Apr. 2011.

\bibitem{harrison2011quantum}
K.~A. Harrison, W.~J. Munro, J.~G. Rarity, and J.~L. Duligall, ``Quantum key
  distribution apparatus and method,'' Nov.~8 2011, uS Patent 8,054,976.

\bibitem{8669813}
M.~{Obeed}, A.~M. {Salhab}, M.-S. Alouini, and S.~A. {Zummo}, ``On optimizing
  {VLC} networks for downlink multi-user transmission: A survey,'' \emph{IEEE
  Communications Surveys Tutorials}, 2019.

\bibitem{6576336}
M.~{Niemiec} and A.~R. {Pach}, ``Management of security in quantum
  cryptography,'' \emph{IEEE Communications Magazine}, vol.~51, no.~8, pp.
  36--41, Aug. 2013.

\bibitem{8425613}
R.~{Henry}, A.~{Herzberg}, and A.~{Kate}, ``Blockchain access privacy:
  Challenges and directions,'' \emph{IEEE Security Privacy}, vol.~16, no.~4,
  pp. 38--45, July 2018.

\bibitem{8368232}
N.~{Van Huynh} \emph{et~al.}, ``Ambient backscatter communications: A
  contemporary survey,'' \emph{IEEE Communications Surveys Tutorials}, vol.~20,
  no.~4, pp. 2889--2922, Fourthquarter 2018.

\bibitem{4600214}
R.~{Madan}, N.~B. {Mehta}, A.~F. {Molisch}, and J.~{Zhang}, ``Energy-efficient
  cooperative relaying over fading channels with simple relay selection,''
  \emph{IEEE Transactions on Wireless Communications}, vol.~7, no.~8, pp.
  3013--3025, Aug. 2008.

\bibitem{7010521}
S.~F. {Yunas}, M.~{Valkama}, and J.~{Niemelä}, ``Spectral and energy
  efficiency of ultra-dense networks under different deployment strategies,''
  \emph{IEEE Communications Magazine}, vol.~53, no.~1, pp. 90--100, Jan. 2015.

\bibitem{7010878}
S.~{Ulukus} \emph{et~al.}, ``Energy harvesting wireless communications: A
  review of recent advances,'' \emph{IEEE Journal on Selected Areas in
  Communications}, vol.~33, no.~3, pp. 360--381, Mar. 2015.

\bibitem{6757202}
J.~L. {Li}, M.~{Krairiksh}, T.~A. {Rahman}, and A.~{Al-Shamma'a}, ``Keynote
  speakers: Wireless power transfer: From long-distance transmission to
  short-range charging,'' in \emph{2013 IEEE International RF and Microwave
  Conference (RFM)}, Dec. 2013, pp. xi--xv.

\bibitem{8382166}
Q.~{Mao}, F.~{Hu}, and Q.~{Hao}, ``Deep learning for intelligent wireless
  networks: A comprehensive survey,'' \emph{IEEE Communications Surveys
  Tutorials}, vol.~20, no.~4, pp. 2595--2621, Fourthquarter 2018.

\bibitem{4078950}
L.~{Yang} and F.~{Wang}, ``Driving into intelligent spaces with pervasive
  communications,'' \emph{IEEE Intelligent Systems}, vol.~22, no.~1, pp.
  12--15, Jan. 2007.

\bibitem{basar2019wireless}
E.~Basar \emph{et~al.}, ``Wireless communications through reconfigurable
  intelligent surfaces,'' \emph{IEEE Access}, 2019.

\bibitem{8493126}
N.~{Javaid}, A.~{Sher}, H.~{Nasir}, and N.~{Guizani}, ``Intelligence in
  {IoT}-based {5G} networks: Opportunities and challenges,'' \emph{IEEE
  Communications Magazine}, vol.~56, no.~10, pp. 94--100, Oct. 2018.

\bibitem{8612930}
A.~{Belmonte-Hernández}, G.~{Hernández-Peñaloza}, D.~{Martín Gutiérrez},
  and F.~{Álvarez}, ``{SWiBluX}: Multi-sensor deep learning fingerprint for
  precise real-time indoor tracking,'' \emph{IEEE Sensors Journal}, vol.~19,
  no.~9, pp. 3473--3486, May 2019.

\bibitem{7156004}
N.~{Zhu} \emph{et~al.}, ``Bridging e-health and the internet of things: The
  {SPHERE} project,'' \emph{IEEE Intelligent Systems}, vol.~30, no.~4, pp.
  39--46, July 2015.

\bibitem{8255764}
M.~{Alzenad}, M.~Z. {Shakir}, H.~{Yanikomeroglu}, and M.-S. Alouini,
  ``{FSO}-based vertical backhaul/fronthaul framework for {5G+} wireless
  networks,'' \emph{IEEE Communications Magazine}, vol.~56, no.~1, pp.
  218--224, Jan. 2018.

\bibitem{kishk2019capacity}
M.~A. Kishk, A.~Bader, and M.-S. Alouini, ``Capacity and coverage enhancement
  using long-endurance tethered airborne base stations,'' \emph{arXiv preprint
  arXiv:1906.11559}, 2019.

\bibitem{8901159}
H.~{Elayan}, O.~{Amin}, B.~{Shihada}, R.~M. {Shubair}, and M.~{Alouini},
  ``Terahertz band: {The} last piece of {RF} spectrum puzzle for communication
  systems,'' \emph{IEEE Open Journal of the Communications Society}, 2019.

\bibitem{sengupta2018terahertz}
K.~Sengupta, T.~Nagatsuma, and D.~M. Mittleman, ``Terahertz integrated
  electronic and hybrid electronic--photonic systems,'' \emph{Nature
  Electronics}, vol.~1, no.~12, p. 622, 2018.

\bibitem{nagatsuma2016advances}
T.~Nagatsuma, G.~Ducournau, and C.~C. Renaud, ``Advances in terahertz
  communications accelerated by photonics,'' \emph{Nature Photonics}, vol.~10,
  no.~6, p. 371, 2016.

\bibitem{drake2006mobile}
F.~Drake, ``Mobile phone masts: protesting the scientific evidence,''
  \emph{Public understanding of science}, vol.~15, no.~4, pp. 387--410, 2006.

\bibitem{6807642}
M.~{Tesanovic} \emph{et~al.}, ``The {LEXNET} project: Wireless networks and
  {EMF}: Paving the way for low-{EMF} networks of the future,'' \emph{IEEE
  Vehicular Technology Magazine}, vol.~9, no.~2, pp. 20--28, Jun. 2014.

\bibitem{pieters2011explanation}
W.~Pieters, ``Explanation and trust: what to tell the user in security and
  {AI}?'' \emph{Ethics and Information Technology}, vol.~13, no.~1, pp. 53--64,
  2011.

\bibitem{8760250}
H.~{Yang} and M.-S. Alouini, ``Data-oriented wireless transmission in future
  wireless systems: Toward trustworthy support of advanced {Internet of
  Things},'' \emph{IEEE Vehicular Technology Magazine}, 2019.

\end{thebibliography}

\textbf{\\Author contributions} Prof. Mohamed-Slim Alouini and Prof. Basem Shihada conceived the work and suggested the outline of the paper. Dr. Shuping Dang and Dr. Osama Amin carried out investigations and wrote the paper.
\textbf{\\ Competing interests} The authors declare no competing interests.

\end{document}